\documentclass[aps,prb,twocolumn,groupedaddress]{revtex4-1}

\usepackage{verbatim}
\usepackage{graphicx}
\begin{document}


\title{Josephson effect in $CeCoIn_5$ microbridges as seen via quantum interferometry}


\author{Oleksandr Foyevtsov}
\email[]{foyevtsov@physik.uni-frankfurt.de}
\author{Fabrizio Porrati}
\author{Michael Huth}
\email[]{michael.huth@physik.uni-frankfurt.de}
\affiliation{Physikalisches Institut, Goethe University, Frankfurt am Main, Germany}


\date{\today}

\begin{abstract}
A superconducting quantum interference device (SQUID) was prepared on a micron-sized single crystal using a selected growth domain of a thin film of $CeCoIn_5$ grown by molecular beam epitaxy. SQUID voltage oscillations of good quality were obtained as well as interference effects stemming from the individual Josephson microbridges. The transport characteristics in the superconducting state exhibited several peculiarities which we ascribe to the periodic motion of vortices in the microbridges. The temperature dependence of the Josephson critical current shows good correspondence to the Ambegaokar-Baratoff relation, expected for the ideal Josephson junction. The results indicate a promising pathway to identify the type of order parameter in $CeCoIn_5$ by means of phase-sensitive measurements on microbridges.
\end{abstract}

\pacs{71.27.+a, 03.75.Lm, 85.25.Dq, 74.50.+r}

\maketitle


\section{Introduction}
Heavy fermion (HF) systems are a class of intermetallic compounds in which electronic correlation effects are particularly pronounced. In HF materials the correlation effects manifest themselves as a strong increase of the effective mass of the charge carriers at low temperatures accompanied with a crossover to the behavior of a Landau Fermi liquid \cite{RevModPhys.73.797}. Hybridization between originally localized $f$-states and the itinerant band states of these mainly $Ce$- and $U$-based compounds is causing this behavior. In special cases, either by chemical substitution or by external stimuli, such as applying pressure or a magnetic field, a sub-class of the HF system can be tuned into a quantum critical state \cite{Gegenwart2008}. 
In this state, characterized by the  shift of a magnetic phase transition to $T=0 K$, magnetic fluctuations exert their influence not only at $T=0 K$ but also at higher temperatures and cause strong deviations from Landau's Fermi liquid behavior. Consequently, these materials are said to behave as non-Fermi liquids. In these non-Fermi liquids the low lying charge excitations cannot be simply described as quasi-particle excitations. Nevertheless, superconducting (SC) condensates can form in the non-Fermi liquid state which is, e.g., the case for $CeCoIn_5$ \cite{PhysRevB.64.134524}. 
The nature of the SC state in this HF material is therefore particularly interesting. Several experiments, such as magnetic field dependent heat conductivity \cite{PhysRevLett.95.067002}, specific heat \cite{PhysRevLett.86.5152} and also point contact spectroscopy \cite{PhysRevB.72.052509} have provided evidence for a $d_{x^2-y^2}$ order parameter. As has been demonstrated for selected high temperature superconductors, phase sensitive experiments employing the Josephson effect can be very helpful in elucidating the order parameter type in a very direct way. This pathway for HF superconductors is, however, difficult to follow. It mainly relies on the availability of epitaxial thin films. 
For selected materials efforts in this direction have been successful \cite{Huth1994116,Huth2001203,Jourdan2004E163} and have eventually led to well characterized tunnel junctions \cite{Jourdan1997335,PlanarUPdAl,zakharov:655}. For $CeCoIn_5$, however, attempts by severals groups have so far not led to epitaxial growth \cite{0953-8984-19-5-056006,Zaitsev200952,izaki:122507}. As a possible way out of this difficulty we have previously developed a technique for addressing individual growth domains or microcrystals in $c$-axis oriented thin films of $CeCoIn_5$ by combining focused ion beam (FIB) etching and focused particle beam induced deposition with appropriate precursors \cite{Foyevtsov20107064,1367-2630-11-3-033032,0957-4484-20-19-195301}. 
In this work we take advantage of the de-wetting tendency during early stages of $CeCoIn_5$ thin film growth on $a$-plane sapphire for selecting an individual growth domain or microcrystal of about two micrometer lateral dimensions suitable for the fabrication of a microbridge-based SQUID. Both, the SQUID loop and the individual microbridges exhibit well-defined voltage modulations in an external magnetic field. The microbridges critical dc Josephson current is in good agreement with the Ambegaokar-Baratoff (AB) model \cite{PhysRevLett.10.486}. Furthermore, in the dynamic resistance we see clear signatures of periodic vortex motion in the microbridges. These results point towards a very promising approach for phase sensitive studies in HF superconductors.

\section{Preparations}
The $CeCoIn_5$ thin film was grown on a $10\times10$mm$^2$ $a-$plane $\alpha-Al_2O_3$ substrate by molecular beam epitaxy method with the growth details and a typical surface morphology reported earlier \cite{Foyevtsov20107064}. The thin film was pre-patterned with conventional ultra violet photolithography and ion beam etching. The micro-structuring (see Fig.~1 (a) and (b)) was performed with FIB milling in an FEI Nova NanoLab 600 dual beam scanning electron microscope (SEM). The outer electrodes (not shown in the figure) were metallic $W$-composite leads prepared by FIB induced deposition using $W(CO)_6$ as precursor gas. A SEM micrograph of the prepared SQUID is shown in Fig.~1c and a detailed view of one of the  microbridges in the inset. The electrical measurements were performed in a $^3He$ cryostat employing a four-probe method and a standard differential resistance measurement technique using current modulation.

\begin{figure}
\includegraphics[width=8cm]{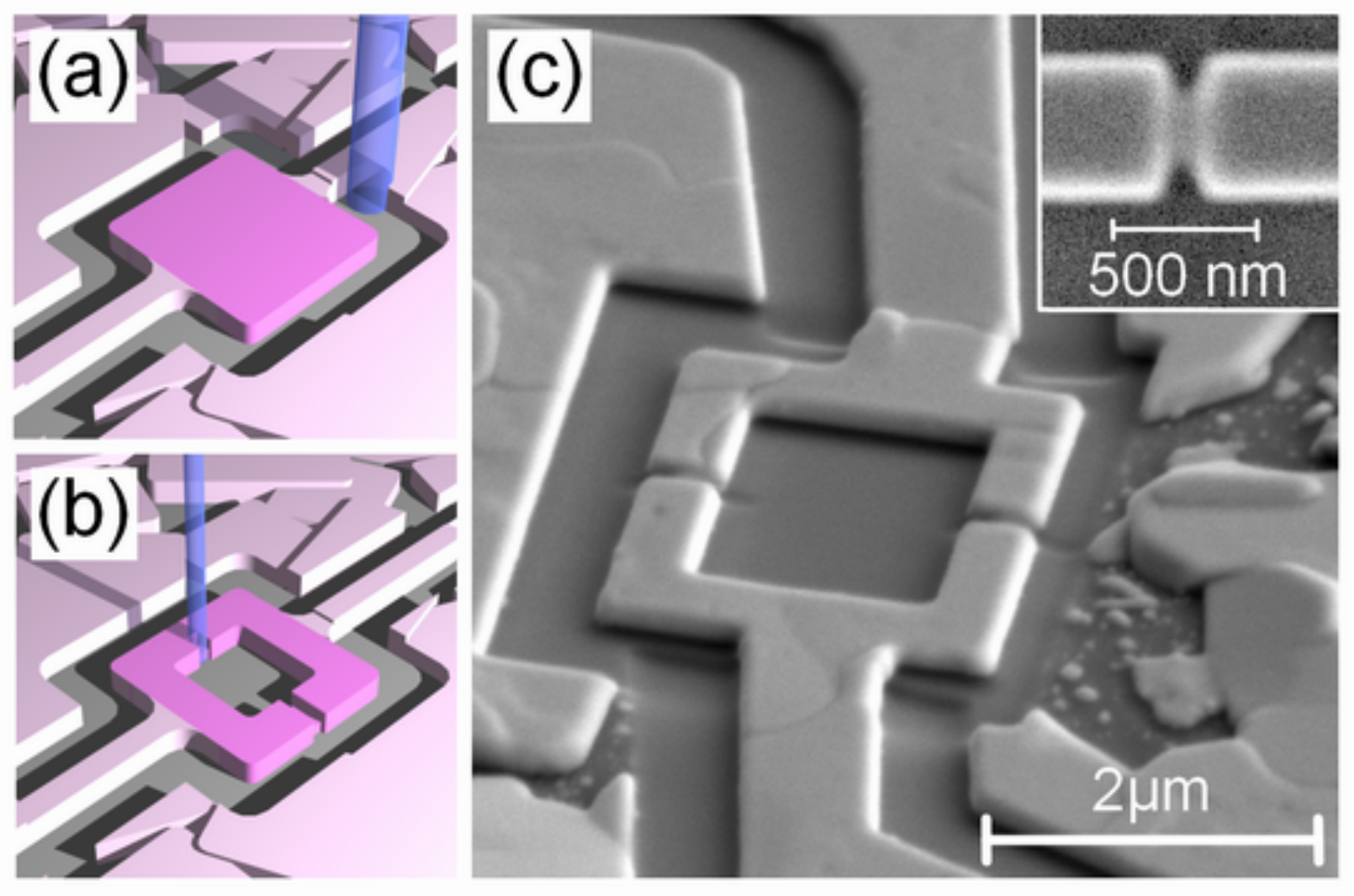}%
\caption{(Color online) (a) FIB milling of the microcrystal with its current and voltage leads out of the interconnected crystals, (b) FIB milling of the loop center and the bridges, (c) SEM micrograph of the such prepared SQUID. The inset in (c) shows a detailed view of one of the two bridges. The loop dimensions are $2.1\mu m\times 1.9\mu m$, as a mean value between the outer and inner diameters, respectively. The estimated microbridges length and width are $L\approx 90nm$ and width $W\approx 250nm$, respectively. The film thickness is about $300nm$.}
\end{figure}

\section{Measurements}
The four-probe resistance of the loop was measured during cool down at a small constant current $I$, and its low temperature region is shown in the inset of Fig.~2. The complete curve shows the typical behavior for $CeCoIn_5$, as was reported for microcrystal measurements before \cite{Foyevtsov20107064}. The residual resistance ratio $R_{300K}/R_{above T_c}$ was found to be 2.9, and is slightly better than in the aforementioned work. However, this value is still small when compared to the value for bulk crystals, which indicates a lower crystal quality for this microcrystal. The resistivity at $2.5 K$ is also slightly larger than was reported for single crystals \cite{0953-8984-13-17-103}. The onset of superconductivity is found at $2.0 K$, somewhat lower than that of a bulk crystal. Another smooth transition is found at around $1.5 K$, which we attribute to the Josephson coupling temperature $T_J$. This transition we ascribe to the SC transition in the bridges, and the local $T_c$ being reduced by the FIB milling process. Also, an important feature is the finite resistance $R_0\approx0.07 \Omega$ (also when $I\to 0$) at the lowest temperature, which we explain as follows. Thermally activated processes may fundamentally modify the voltage-to-current $V-I$ relation of a Josephson junction. In particular, as was shown by Ambegaokar and Halperin \cite{PhysRevLett.22.1364}, there is always a finite resistance even below the Josephson critical current $I_c$. This may be viewed as an overdamped junction, i.e. a junction for which the McCumber parameter is small, which is expected to be the case for microbridges \cite{RevModPhys.51.101}. 

The theoretical analysis of bridges whose dimensions are large compared to the Ginsburg-Landau (GL) coherence length $\xi$ is difficult. For large ratios of $W/\xi$ and $L/\xi$ the bridge leaves the regime of the 'ideal' Josephson behavior and enters the Abricosov vortex motion regime, as was pointed out by Likharev \cite{RevModPhys.51.101}. As a result, essential characteristics of the bridge are modified, such as the shape of SQUID voltage modulations or the dependence $I_c(T)$ \cite{RevModPhys.51.101,RevModPhys.76.411}. The dimensions of our bridges when compared with $\xi_{CeCoIn_5}\approx 5 nm$ \cite{PhysRevLett.97.127001} make both ratios $W/\xi\approx 18$ and $L/\xi\approx 50$ large. 

\begin{figure}
\includegraphics[width=8cm]{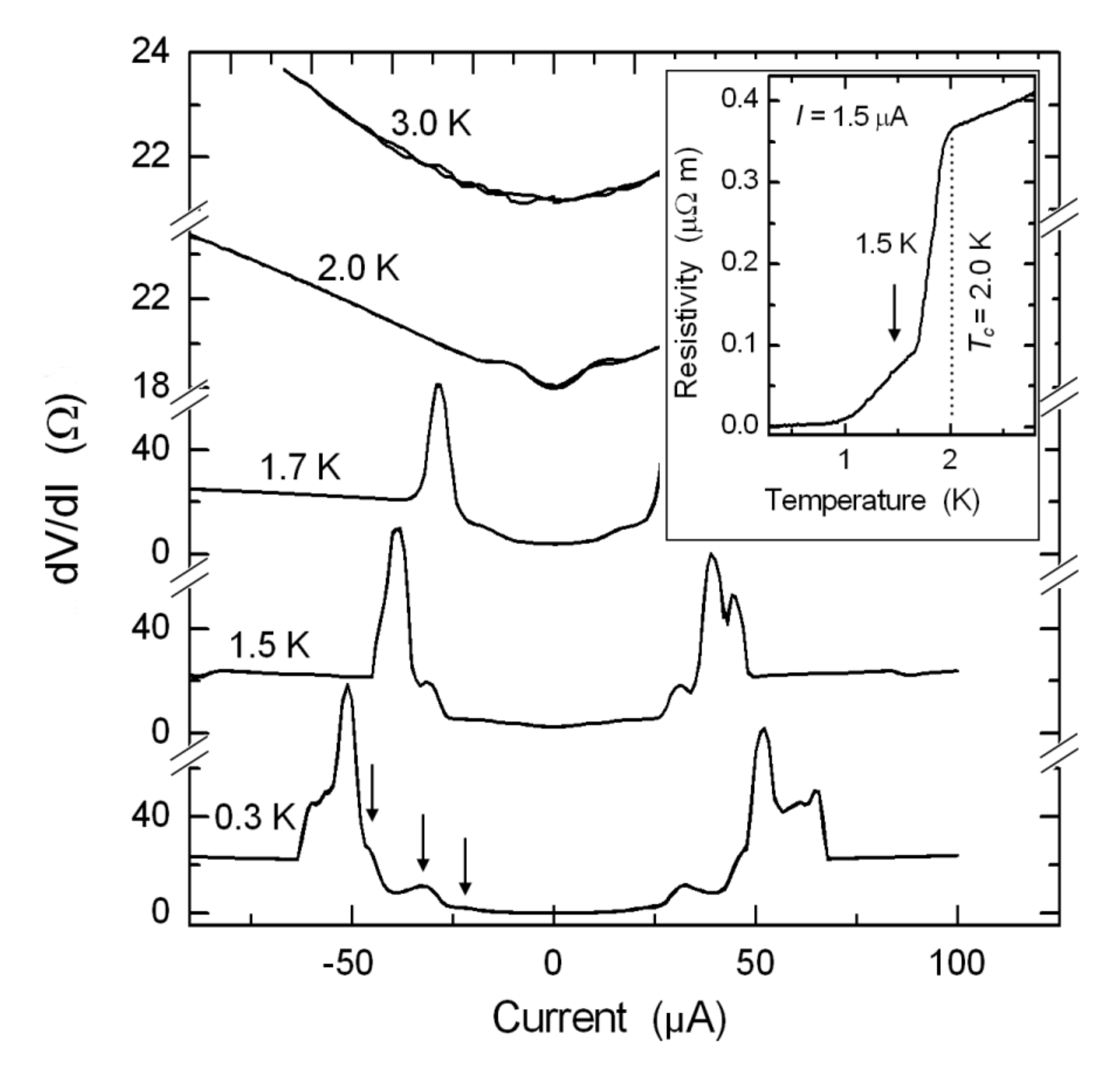}%
\caption{Set of measured dynamic resistance curves for the SQUID as a function of current. The features on the lowest curve are ascribed to the periodic motion of vortices and shown with arrows (see text for details). The inset shows the measured four-probe resistance during cool down in the low-T region. The arrow denotes an additional transition at $1.5 K$ associated to the Josephson coupling temperature $T_J$. Note the non-zero resistivity even at the lowest temperatures.}
\end{figure}

In the main part of Fig.~2 we show a set of measured dynamic resistance curves as a function of current plotted for selected temperatures. The curves measured above $T_c$ show an anomalous background resistance, visible up to $10 K$, out of which evolves the SC state. This nonlinearity, which we do not attribute to self heating effects, will be addressed in a separate publication. Signatures of superconductivity become evident on top of the background anomaly at about $2.0 K$ and become more pronounced as the temperature decreases. At temperatures below $T_J$ the curves develop additional features, which we marked by arrows on the lowest curve.

The inset in Fig.~3 shows typical measured SQUID voltage oscillations as a function of an externally applied magnetic field. The oscillations are well reproducible and of sine-like shape without asymmetries. The period of the measured oscillations corresponds to one flux quantum $\Phi_0=h/2e$ over the averaged loop area with excellent agreement. Occasionally, the oscillations become phase-shifted by a random amount. Since the oscillations are a measure of the SC phase difference between the two SC arms modulated by an external magnetic field, we attribute these phase slips to randomly trapped flux.
The main part of Fig.~3 shows a long period modulation superimposed with the SQUID modulations, which appear as a band due to the scale. We ascribe them to the Josephson modulations of the two individual bridges. These oscillations are also affected by random phase slips. Their perturbed periodicity might be caused by interference effects between both bridges, since they are coupled by the SC wave function in the arms. A rough estimate of the period for these oscillations gives $~10 mT$, and according to $s=\Phi_0/B$ implies that the effective dimensions of each bridge extend into each SC bank by about $200 nm$, which compares favorably with the London penetration depth $\lambda_L$ for $CeCoIn_5$ \cite{PhysRevLett.97.127001}.

\begin{figure}
\includegraphics[width=8cm]{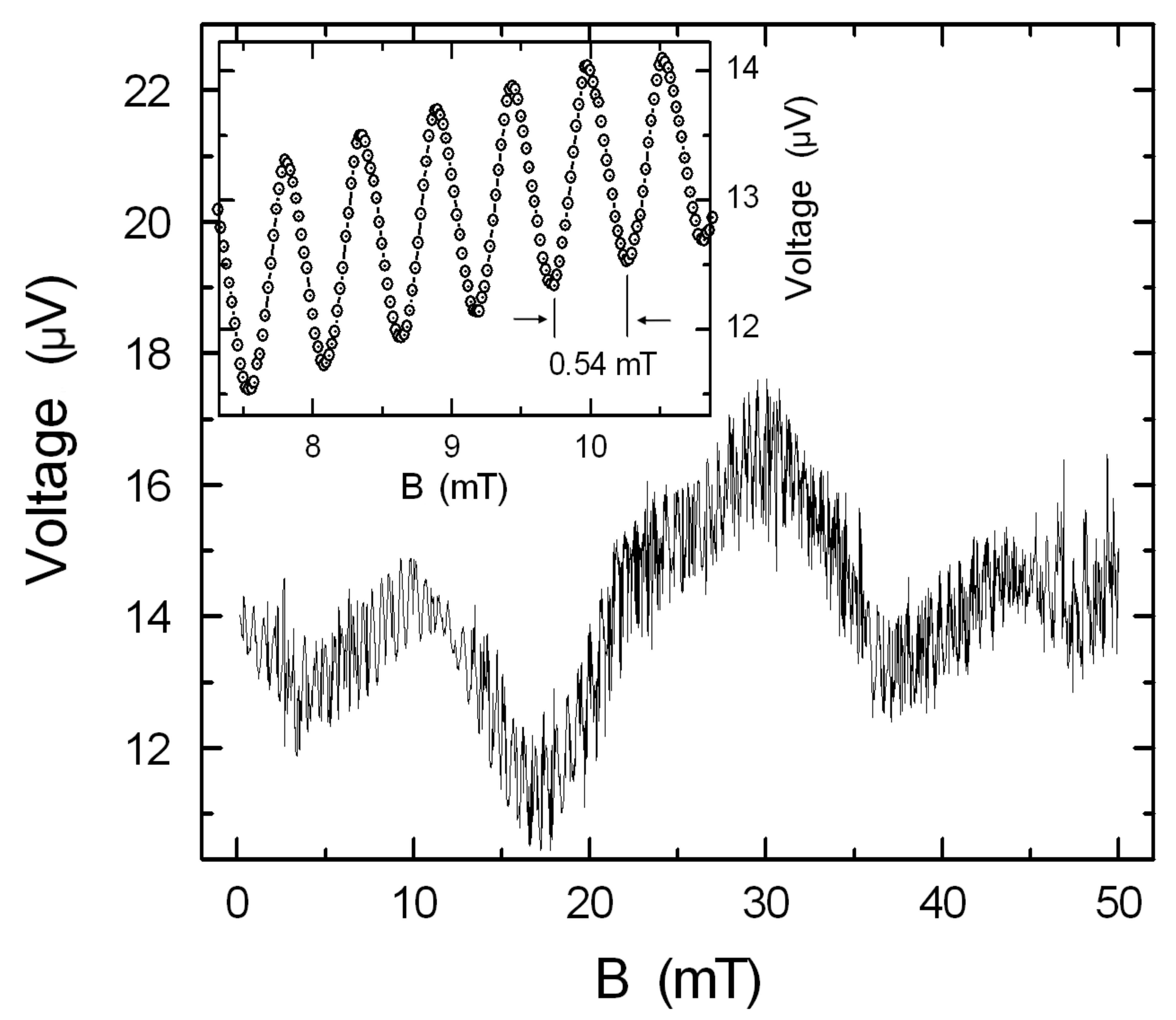}%
\caption{Measured voltage oscillations as a function of external magnetic field on a large scale. The inset shows the small scale SQUID oscillations. Note in the main part that due to the scale the SQUID oscillations appear as a band of values on top of the slow modulation.}
\end{figure}

\section{Analysis}
The observed strong manifestation of the Josephson effect enables us to analyze the peculiarities of the transport characteristics at low temperatures in the framework of models developed for the Josephson effect in microbridges \cite{AL.1975,KO.1975,KO.1977,PTP.44.1525}. The Aslamazov-Larkin (AL) theory \cite{AL.1975} may be used for the analysis of the $V-I$ characteristics of microbridges of a geometry similar to ours when vortex dynamics has to be accounted for, i.e. $W\gg\xi$ and $L>\xi$. In the framework of this model the electrical current density across the bridge is non-uniform, and is described as $j(x)=J/\pi d\sqrt{a^2-x^2}$, where $J$ is the total current along the bridge, $d$ is the bridge thickness, $a$ is equal to $W/2$, and $x$ is the coordinate along the width of the bridge with $x=0$ at the bridge center. It may be shown that two vortices of opposite helicity are most likely to enter the bridge from opposite edges due to the locally highest current density there. 

As is shown in the AL model, the total force acting on the $n-th$ vortex is $F_n=-F_L+F_s\pm F_{vv}$. Here, $F_L$ is the Lorentz force due to the interaction with current $J$, $F_s$ is the interaction with the edge, and $F_{vv}$ is the vortex-vortex interaction term. At low T a vortex remains pinned to the edge due to $F_s$ as long as the counteracting Lorentz force is $F_L\le F_s$. Neglecting the $F_{vv}$ term when the number of vortices in the bridge is small, one may derive the current at which $F_L=F_s$ as
\begin{math}
 J_C\approx J_0\sqrt(a/2\xi)
\end{math},
where $J_0=c^2\hbar d/8e\lambda_L^2$. At this and larger currents vortex pairs start a periodic viscous motion toward the bridge center with successive annihilation and new creation of a vortex pair at the edges. As the $J$ and $F_L$ increase, more than one vortex pair may be present simultaneously within the bridge giving a discontinuity on the $V-I$ curve. However, due to the associated increase of the $F_{vv}$, any succeeding current span will differ from the previous one. The equation of motion of the $n-th$ vortex is given by $\dot{x}_n=F_n/\eta$, where $\eta$ is the viscosity coefficient. Finally, solving this equation one finds the period of moving vortices, which is equal to the period $T$ in the Josephson frequency relation $\hbar \omega=2\pi\hbar/T=2e\bar{V}$, where $\bar{V}$ is the averaged measured voltage. Hence, the electrical current $J$ is related with $\bar{V}$. According to the AL model discontinuities on the $V-I$ curves are periodic in the voltage. One may show that this periodicity is given by (SI units):

\begin{equation}
 V_0=\frac{1}{2\sqrt{2}} \frac{4\pi}{\mu_0} \frac{d\hbar^3}{a\sqrt{\xi a}\lambda_{eff}^2} \frac{1}{e^3\eta}
\end{equation}

where $\lambda_{eff}$ is the effective penetration depth, which we take as $\lambda_L$, since $\lambda_L\approx d$ and the correction is small.
We attribute the smoothed peaks marked with arrows in Fig.~2 each to a new number of vortex pairs moving simultaneously in the bridge. They appear, as expected, just above $I_c$. Three corresponding peaks are well resolved, which in our case implies that the strong repulsion between neighbored vortices of equal helicity limits considerably their number. This is very probable because the average inter-vortex distance $a/3$ is comparable with the vortex radius. Also, large values of $\lambda_L\approx 235 nm$ \cite{0295-5075-62-3-412} for $CeCoIn_5$ do increase the inter-vortex repulsive coupling.
The peaks are not equidistant with current, which is also expected from the AL model, but the voltage peaks are equidistant with a period of about $125 \mu V$ with good accuracy. This periodicity we ascribe to $V_0$ as predicted by the AL model. One may obtain the viscosity coefficient $\eta$ from Eq.~(1). Using $\lambda_L=235nm$, the approximate thickness in the bridge as $d/2 \approx 150 nm$, and $\xi=10nm$ \cite{PhysRevLett.97.127001} we find $\eta\approx5.0\times10^{-15} (AVs^2m^{-2})$. The agreement with the Bardeen-Stephen theory \cite{PhysRev.140.A1197} is good, which gives $\eta\approx 2.6\times10^{-15} (AVs^2m^{-2})$ using $\xi=10 nm$ and the normal state resistivity $\rho_n=4.0\times 10^{-7} \Omega m$.

\begin{figure}
\includegraphics[width=7.5cm]{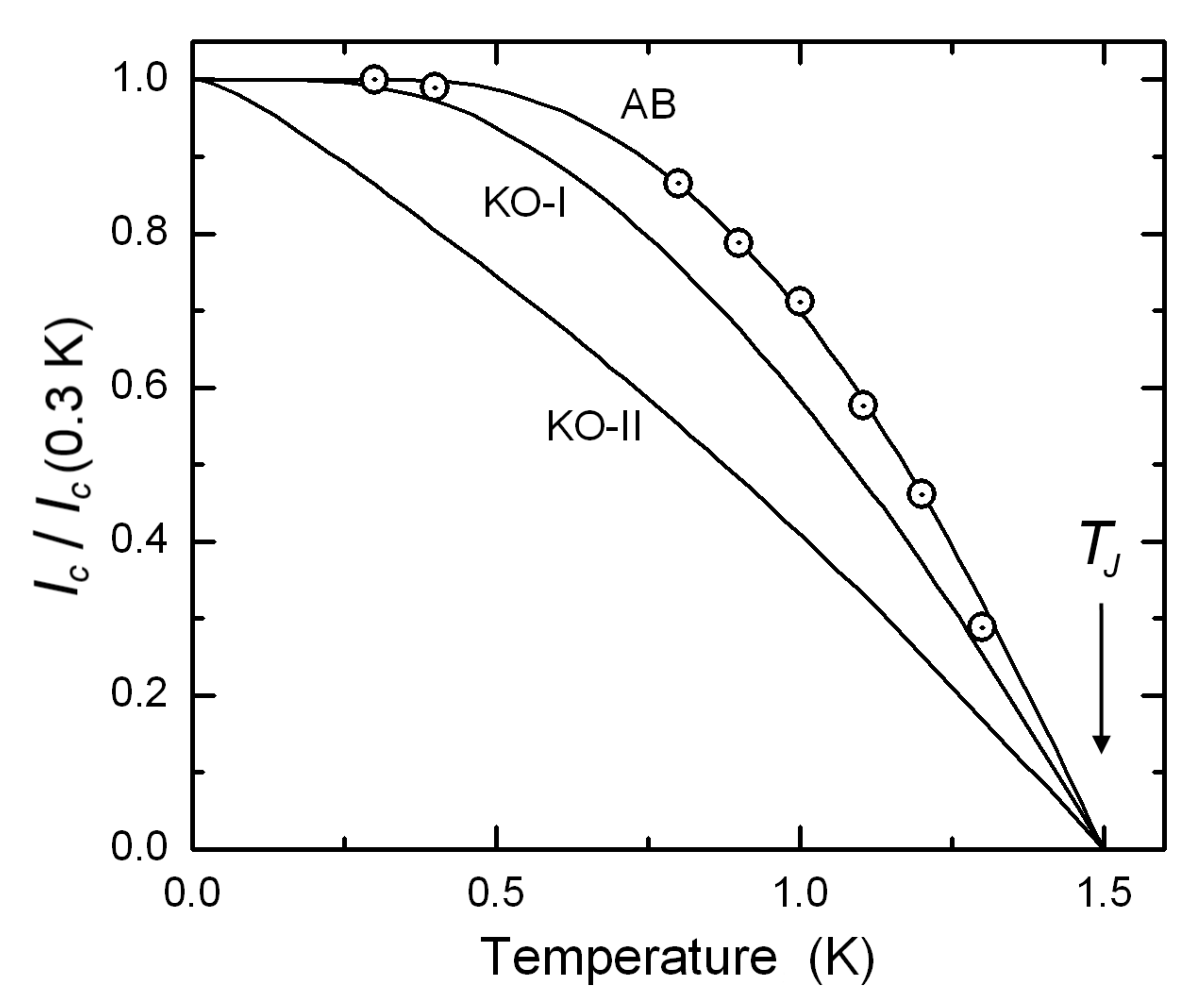}%
\caption{The reduced Josephson critical current for an individual microbridge plotted versus temperature. Circles represent the measured data. Solid curves are calculated for: (AB)-Ambegaokar-Baratoff, Kulik-Omel'yanchuk models for diffusive regime (KO-I), and for ballistic regime (KO-II). $T_J$ denote the Josephson coupling temperature. The measured value of the normal state resistance $R_n\approx42\Omega$, while effective one used for the AB fit is $68\Omega$. The measured $I_c$ at the lowest temperature is $\approx5.2\mu A$}
\end{figure}

A temperature dependence of the reduced Josephson critical current $I_c$ for the two bridges is shown in Fig.~4. The dots represent measured data and the solid lines are theoretical curves for several cases using the measured $T_J$ and the normal state resistance $R_n$. It is clear that the AB model \cite{PhysRevLett.10.486} fits the data well, although $R_n = 68 \Omega$ has to be used instead of the measured value of 42$\Omega$. This is not unexpected. On the one hand, the AB fit tends to overestimate the value of $I_c$ for $L>\xi$, since $R_n$ growth linearly with $L$ but the GL order parameter in the middle of the bridge falls off exponentially. On the other hand, the AB model is derived for tunnel junctions, that is $L\ll\xi$ and with no scattering within the barrier. In contradistinction, models derived for microconstrictions, such as that by Kulik and Omel'yanchuk, in the dirty and clean limit \cite{KO.1975,KO.1977} or Ishii's model \cite{PTP.44.1525} for long bridges do not fit the data. Additionally, the shape of the voltage modulations are not asymmetric, as expected in these models. We have to state that the observed $I_c(T)$ dependence does not follow the probably expected behavior theoretically predicted for microbridges in the dirty or clean limit. These models do not take a possible influence of Andreev bound states \cite{PhysRevB.76.064529} for a $d$-wave order parameter into account, which may be one reason for this discrepancy.

\section{Conclusions}
The clean SQUID characteristics observed in this work on selected $CeCoIn_5$ microcrystals allow for two main conclusions. First, the fact that the micro-patterning approach used here is not detrimental to the superconducting state in $CeCoIn_5$ indicates a certain robustness of this state. Considering, as a rule, the high sensitivity of the superconducting properties in other heavy-fermion superconductors, this may point toward an enhanced  stability of the superconducting state when formed from a non-Fermi liquid. Second, the approach followed here provides junctions with rather close-to-ideal characteristics, which is promising with regard to utilizing other microbridge geometries. These geometries, such as recently proposed by Gumann and Schopohl \cite{PhysRevB.79.144505} or also a $\pi$-junction structure, which can e.g. be formed by the combination of a conventional SC lead prepared by focused ion beam induced deposition using the $W(CO)_6$ precursor with a $CeCoIn_5$ microcrystal, can provide phase-sensitive information about the superconducting state in $CeCoIn_5$. Work along these lines is in progress.

\section*{Acknowledgments}
We would like to acknowledge the Deutsche Forschungsgemeinschaft (DFG) for financial support under grant no. HU 752/3-3 and Martin Jourdan for granting access to an ion beam etching facility.


\begin{thebibliography}{32}%
\makeatletter
\providecommand \@ifxundefined [1]{%
 \@ifx{#1\undefined}
}%
\providecommand \@ifnum [1]{%
 \ifnum #1\expandafter \@firstoftwo
 \else \expandafter \@secondoftwo
 \fi
}%
\providecommand \@ifx [1]{%
 \ifx #1\expandafter \@firstoftwo
 \else \expandafter \@secondoftwo
 \fi
}%
\providecommand \natexlab [1]{#1}%
\providecommand \enquote  [1]{``#1''}%
\providecommand \bibnamefont  [1]{#1}%
\providecommand \bibfnamefont [1]{#1}%
\providecommand \citenamefont [1]{#1}%
\providecommand \href@noop [0]{\@secondoftwo}%
\providecommand \href [0]{\begingroup \@sanitize@url \@href}%
\providecommand \@href[1]{\@@startlink{#1}\@@href}%
\providecommand \@@href[1]{\endgroup#1\@@endlink}%
\providecommand \@sanitize@url [0]{\catcode `\\12\catcode `\$12\catcode
  `\&12\catcode `\#12\catcode `\^12\catcode `\_12\catcode `\%12\relax}%
\providecommand \@@startlink[1]{}%
\providecommand \@@endlink[0]{}%
\providecommand \url  [0]{\begingroup\@sanitize@url \@url }%
\providecommand \@url [1]{\endgroup\@href {#1}{\urlprefix }}%
\providecommand \urlprefix  [0]{URL }%
\providecommand \Eprint [0]{\href }%
\providecommand \doibase [0]{http://dx.doi.org/}%
\providecommand \selectlanguage [0]{\@gobble}%
\providecommand \bibinfo  [0]{\@secondoftwo}%
\providecommand \bibfield  [0]{\@secondoftwo}%
\providecommand \translation [1]{[#1]}%
\providecommand \BibitemOpen [0]{}%
\providecommand \bibitemStop [0]{}%
\providecommand \bibitemNoStop [0]{.\EOS\space}%
\providecommand \EOS [0]{\spacefactor3000\relax}%
\providecommand \BibitemShut  [1]{\csname bibitem#1\endcsname}%
\let\auto@bib@innerbib\@empty
\bibitem [{\citenamefont {Stewart}(2001)}]{RevModPhys.73.797}%
  \BibitemOpen
  \bibfield  {author} {\bibinfo {author} {\bibfnamefont {G.~R.}\ \bibnamefont
  {Stewart}},\ }\href {\doibase 10.1103/RevModPhys.73.797} {\bibfield
  {journal} {\bibinfo  {journal} {Rev. Mod. Phys.}\ }\textbf {\bibinfo {volume}
  {73}},\ \bibinfo {pages} {797} (\bibinfo {year} {2001})}\BibitemShut
  {NoStop}%
\bibitem [{\citenamefont {Gegenwart}\ \emph {et~al.}(2008)\citenamefont
  {Gegenwart}, \citenamefont {Si},\ and\ \citenamefont
  {Steglich}}]{Gegenwart2008}%
  \BibitemOpen
  \bibfield  {author} {\bibinfo {author} {\bibfnamefont {P.}~\bibnamefont
  {Gegenwart}}, \bibinfo {author} {\bibfnamefont {Q.}~\bibnamefont {Si}}, \
  and\ \bibinfo {author} {\bibfnamefont {F.}~\bibnamefont {Steglich}},\ }\href
  {\doibase 10.1126/science.1191195} {\bibfield  {journal} {\bibinfo  {journal}
  {Nature Physics}\ }\textbf {\bibinfo {volume} {4}},\ \bibinfo {pages} {186}
  (\bibinfo {year} {2008})}\BibitemShut {NoStop}%
\bibitem [{\citenamefont {Kim}\ \emph {et~al.}(2001)\citenamefont {Kim},
  \citenamefont {Alwood}, \citenamefont {Stewart}, \citenamefont {Sarrao},\
  and\ \citenamefont {Thompson}}]{PhysRevB.64.134524}%
  \BibitemOpen
  \bibfield  {author} {\bibinfo {author} {\bibfnamefont {J.~S.}\ \bibnamefont
  {Kim}}, \bibinfo {author} {\bibfnamefont {J.}~\bibnamefont {Alwood}},
  \bibinfo {author} {\bibfnamefont {G.~R.}\ \bibnamefont {Stewart}}, \bibinfo
  {author} {\bibfnamefont {J.~L.}\ \bibnamefont {Sarrao}}, \ and\ \bibinfo
  {author} {\bibfnamefont {J.~D.}\ \bibnamefont {Thompson}},\ }\href {\doibase
  10.1103/PhysRevB.64.134524} {\bibfield  {journal} {\bibinfo  {journal} {Phys.
  Rev. B}\ }\textbf {\bibinfo {volume} {64}},\ \bibinfo {pages} {134524}
  (\bibinfo {year} {2001})}\BibitemShut {NoStop}%
\bibitem [{\citenamefont {Tanatar}\ \emph {et~al.}(2005)\citenamefont
  {Tanatar}, \citenamefont {Paglione}, \citenamefont {Nakatsuji}, \citenamefont
  {Hawthorn}, \citenamefont {Boaknin}, \citenamefont {Hill}, \citenamefont
  {Ronning}, \citenamefont {Sutherland}, \citenamefont {Taillefer},
  \citenamefont {Petrovic}, \citenamefont {Canfield},\ and\ \citenamefont
  {Fisk}}]{PhysRevLett.95.067002}%
  \BibitemOpen
  \bibfield  {author} {\bibinfo {author} {\bibfnamefont {M.~A.}\ \bibnamefont
  {Tanatar}}, \bibinfo {author} {\bibfnamefont {J.}~\bibnamefont {Paglione}},
  \bibinfo {author} {\bibfnamefont {S.}~\bibnamefont {Nakatsuji}}, \bibinfo
  {author} {\bibfnamefont {D.~G.}\ \bibnamefont {Hawthorn}}, \bibinfo {author}
  {\bibfnamefont {E.}~\bibnamefont {Boaknin}}, \bibinfo {author} {\bibfnamefont
  {R.~W.}\ \bibnamefont {Hill}}, \bibinfo {author} {\bibfnamefont
  {F.}~\bibnamefont {Ronning}}, \bibinfo {author} {\bibfnamefont
  {M.}~\bibnamefont {Sutherland}}, \bibinfo {author} {\bibfnamefont
  {L.}~\bibnamefont {Taillefer}}, \bibinfo {author} {\bibfnamefont
  {C.}~\bibnamefont {Petrovic}}, \bibinfo {author} {\bibfnamefont {P.~C.}\
  \bibnamefont {Canfield}}, \ and\ \bibinfo {author} {\bibfnamefont
  {Z.}~\bibnamefont {Fisk}},\ }\href {\doibase 10.1103/PhysRevLett.95.067002}
  {\bibfield  {journal} {\bibinfo  {journal} {Phys. Rev. Lett.}\ }\textbf
  {\bibinfo {volume} {95}},\ \bibinfo {pages} {067002} (\bibinfo {year}
  {2005})}\BibitemShut {NoStop}%
\bibitem [{\citenamefont {Movshovich}\ \emph {et~al.}(2001)\citenamefont
  {Movshovich}, \citenamefont {Jaime}, \citenamefont {Thompson}, \citenamefont
  {Petrovic}, \citenamefont {Fisk}, \citenamefont {Pagliuso},\ and\
  \citenamefont {Sarrao}}]{PhysRevLett.86.5152}%
  \BibitemOpen
  \bibfield  {author} {\bibinfo {author} {\bibfnamefont {R.}~\bibnamefont
  {Movshovich}}, \bibinfo {author} {\bibfnamefont {M.}~\bibnamefont {Jaime}},
  \bibinfo {author} {\bibfnamefont {J.~D.}\ \bibnamefont {Thompson}}, \bibinfo
  {author} {\bibfnamefont {C.}~\bibnamefont {Petrovic}}, \bibinfo {author}
  {\bibfnamefont {Z.}~\bibnamefont {Fisk}}, \bibinfo {author} {\bibfnamefont
  {P.~G.}\ \bibnamefont {Pagliuso}}, \ and\ \bibinfo {author} {\bibfnamefont
  {J.~L.}\ \bibnamefont {Sarrao}},\ }\href {\doibase
  10.1103/PhysRevLett.86.5152} {\bibfield  {journal} {\bibinfo  {journal}
  {Phys. Rev. Lett.}\ }\textbf {\bibinfo {volume} {86}},\ \bibinfo {pages}
  {5152} (\bibinfo {year} {2001})}\BibitemShut {NoStop}%
\bibitem [{\citenamefont {Park}\ \emph {et~al.}(2005)\citenamefont {Park},
  \citenamefont {Greene}, \citenamefont {Sarrao},\ and\ \citenamefont
  {Thompson}}]{PhysRevB.72.052509}%
  \BibitemOpen
  \bibfield  {author} {\bibinfo {author} {\bibfnamefont {W.~K.}\ \bibnamefont
  {Park}}, \bibinfo {author} {\bibfnamefont {L.~H.}\ \bibnamefont {Greene}},
  \bibinfo {author} {\bibfnamefont {J.~L.}\ \bibnamefont {Sarrao}}, \ and\
  \bibinfo {author} {\bibfnamefont {J.~D.}\ \bibnamefont {Thompson}},\ }\href
  {\doibase 10.1103/PhysRevB.72.052509} {\bibfield  {journal} {\bibinfo
  {journal} {Phys. Rev. B}\ }\textbf {\bibinfo {volume} {72}},\ \bibinfo
  {pages} {052509} (\bibinfo {year} {2005})}\BibitemShut {NoStop}%
\bibitem [{\citenamefont {Huth}\ \emph {et~al.}(1994)\citenamefont {Huth},
  \citenamefont {Kaldowski}, \citenamefont {Hessert}, \citenamefont {Heske},\
  and\ \citenamefont {Adrian}}]{Huth1994116}%
  \BibitemOpen
  \bibfield  {author} {\bibinfo {author} {\bibfnamefont {M.}~\bibnamefont
  {Huth}}, \bibinfo {author} {\bibfnamefont {A.}~\bibnamefont {Kaldowski}},
  \bibinfo {author} {\bibfnamefont {J.}~\bibnamefont {Hessert}}, \bibinfo
  {author} {\bibfnamefont {C.}~\bibnamefont {Heske}}, \ and\ \bibinfo {author}
  {\bibfnamefont {H.}~\bibnamefont {Adrian}},\ }\href {\doibase DOI:
  10.1016/0921-4526(94)91753-1} {\bibfield  {journal} {\bibinfo  {journal}
  {Physica B: Condensed Matter}\ }\textbf {\bibinfo {volume} {199-200}},\
  \bibinfo {pages} {116 } (\bibinfo {year} {1994})},\ \bibinfo {note} {and
  references therein}\BibitemShut {NoStop}%
\bibitem [{\citenamefont {Huth}\ \emph {et~al.}(2001)\citenamefont {Huth},
  \citenamefont {Meffert}, \citenamefont {Oster},\ and\ \citenamefont
  {Adrian}}]{Huth2001203}%
  \BibitemOpen
  \bibfield  {author} {\bibinfo {author} {\bibfnamefont {M.}~\bibnamefont
  {Huth}}, \bibinfo {author} {\bibfnamefont {H.}~\bibnamefont {Meffert}},
  \bibinfo {author} {\bibfnamefont {J.}~\bibnamefont {Oster}}, \ and\ \bibinfo
  {author} {\bibfnamefont {H.}~\bibnamefont {Adrian}},\ }\href {\doibase DOI:
  10.1016/S0022-0248(01)01476-2} {\bibfield  {journal} {\bibinfo  {journal}
  {Journal of Crystal Growth}\ }\textbf {\bibinfo {volume} {231}},\ \bibinfo
  {pages} {203 } (\bibinfo {year} {2001})}\BibitemShut {NoStop}%
\bibitem [{\citenamefont {Jourdan}\ \emph {et~al.}(2004)\citenamefont
  {Jourdan}, \citenamefont {Zakharov}, \citenamefont {Foerster},\ and\
  \citenamefont {Adrian}}]{Jourdan2004E163}%
  \BibitemOpen
  \bibfield  {author} {\bibinfo {author} {\bibfnamefont {M.}~\bibnamefont
  {Jourdan}}, \bibinfo {author} {\bibfnamefont {A.}~\bibnamefont {Zakharov}},
  \bibinfo {author} {\bibfnamefont {M.}~\bibnamefont {Foerster}}, \ and\
  \bibinfo {author} {\bibfnamefont {H.}~\bibnamefont {Adrian}},\ }\href
  {\doibase DOI: 10.1016/j.jmmm.2003.11.095} {\bibfield  {journal} {\bibinfo
  {journal} {Journal of Magnetism and Magnetic Materials}\ }\textbf {\bibinfo
  {volume} {272-276}},\ \bibinfo {pages} {E163 } (\bibinfo {year}
  {2004})}\BibitemShut {NoStop}%
\bibitem [{\citenamefont {Jourdan}\ \emph {et~al.}(1997)\citenamefont
  {Jourdan}, \citenamefont {Huth}, \citenamefont {Hessert},\ and\ \citenamefont
  {Adrian}}]{Jourdan1997335}%
  \BibitemOpen
  \bibfield  {author} {\bibinfo {author} {\bibfnamefont {M.}~\bibnamefont
  {Jourdan}}, \bibinfo {author} {\bibfnamefont {M.}~\bibnamefont {Huth}},
  \bibinfo {author} {\bibfnamefont {J.}~\bibnamefont {Hessert}}, \ and\
  \bibinfo {author} {\bibfnamefont {H.}~\bibnamefont {Adrian}},\ }\href
  {\doibase DOI: 10.1016/S0921-4526(96)00710-7} {\bibfield  {journal} {\bibinfo
   {journal} {Physica B: Condensed Matter}\ }\textbf {\bibinfo {volume}
  {230-232}},\ \bibinfo {pages} {335 } (\bibinfo {year} {1997})}\BibitemShut
  {NoStop}%
\bibitem [{\citenamefont {Jourdan}\ \emph {et~al.}(1999)\citenamefont
  {Jourdan}, \citenamefont {Huth},\ and\ \citenamefont {Adrian}}]{PlanarUPdAl}%
  \BibitemOpen
  \bibfield  {author} {\bibinfo {author} {\bibfnamefont {M.}~\bibnamefont
  {Jourdan}}, \bibinfo {author} {\bibfnamefont {M.}~\bibnamefont {Huth}}, \
  and\ \bibinfo {author} {\bibfnamefont {H.}~\bibnamefont {Adrian}},\
  }\href@noop {} {\bibfield  {journal} {\bibinfo  {journal} {Nature}\ }\textbf
  {\bibinfo {volume} {398}},\ \bibinfo {pages} {47} (\bibinfo {year}
  {1999})}\BibitemShut {NoStop}%
\bibitem [{\citenamefont {Zakharov}\ \emph {et~al.}(2006)\citenamefont
  {Zakharov}, \citenamefont {Jourdan},\ and\ \citenamefont
  {Adrian}}]{zakharov:655}%
  \BibitemOpen
  \bibfield  {author} {\bibinfo {author} {\bibfnamefont {A.}~\bibnamefont
  {Zakharov}}, \bibinfo {author} {\bibfnamefont {M.}~\bibnamefont {Jourdan}}, \
  and\ \bibinfo {author} {\bibfnamefont {H.}~\bibnamefont {Adrian}},\ }\href
  {\doibase 10.1063/1.2354880} {\bibfield  {journal} {\bibinfo  {journal} {AIP
  Conference Proceedings}\ }\textbf {\bibinfo {volume} {850}},\ \bibinfo
  {pages} {655} (\bibinfo {year} {2006})}\BibitemShut {NoStop}%
\bibitem [{\citenamefont {Soroka}\ \emph {et~al.}(2007)\citenamefont {Soroka},
  \citenamefont {Blendin},\ and\ \citenamefont {Huth}}]{0953-8984-19-5-056006}%
  \BibitemOpen
  \bibfield  {author} {\bibinfo {author} {\bibfnamefont {O.~K.}\ \bibnamefont
  {Soroka}}, \bibinfo {author} {\bibfnamefont {G.}~\bibnamefont {Blendin}}, \
  and\ \bibinfo {author} {\bibfnamefont {M.}~\bibnamefont {Huth}},\ }\href
  {http://stacks.iop.org/0953-8984/19/i=5/a=056006} {\bibfield  {journal}
  {\bibinfo  {journal} {Journal of Physics: Condensed Matter}\ }\textbf
  {\bibinfo {volume} {19}},\ \bibinfo {pages} {056006} (\bibinfo {year}
  {2007})}\BibitemShut {NoStop}%
\bibitem [{\citenamefont {Zaitsev}\ \emph {et~al.}(2009)\citenamefont
  {Zaitsev}, \citenamefont {Beck}, \citenamefont {Schneider}, \citenamefont
  {Fromknecht}, \citenamefont {Fuchs}, \citenamefont {Geerk},\ and\
  \citenamefont {v.~L\"{o}hneysen}}]{Zaitsev200952}%
  \BibitemOpen
  \bibfield  {author} {\bibinfo {author} {\bibfnamefont {A.}~\bibnamefont
  {Zaitsev}}, \bibinfo {author} {\bibfnamefont {A.}~\bibnamefont {Beck}},
  \bibinfo {author} {\bibfnamefont {R.}~\bibnamefont {Schneider}}, \bibinfo
  {author} {\bibfnamefont {R.}~\bibnamefont {Fromknecht}}, \bibinfo {author}
  {\bibfnamefont {D.}~\bibnamefont {Fuchs}}, \bibinfo {author} {\bibfnamefont
  {J.}~\bibnamefont {Geerk}}, \ and\ \bibinfo {author} {\bibfnamefont
  {H.}~\bibnamefont {v.~L\"{o}hneysen}},\ }\href {\doibase DOI:
  10.1016/j.physc.2008.11.001} {\bibfield  {journal} {\bibinfo  {journal}
  {Physica C: Superconductivity}\ }\textbf {\bibinfo {volume} {469}},\ \bibinfo
  {pages} {52 } (\bibinfo {year} {2009})}\BibitemShut {NoStop}%
\bibitem [{\citenamefont {Izaki}\ \emph {et~al.}(2007)\citenamefont {Izaki},
  \citenamefont {Shishido}, \citenamefont {Kato}, \citenamefont {Shibauchi},
  \citenamefont {Matsuda},\ and\ \citenamefont {Terashima}}]{izaki:122507}%
  \BibitemOpen
  \bibfield  {author} {\bibinfo {author} {\bibfnamefont {M.}~\bibnamefont
  {Izaki}}, \bibinfo {author} {\bibfnamefont {H.}~\bibnamefont {Shishido}},
  \bibinfo {author} {\bibfnamefont {T.}~\bibnamefont {Kato}}, \bibinfo {author}
  {\bibfnamefont {T.}~\bibnamefont {Shibauchi}}, \bibinfo {author}
  {\bibfnamefont {Y.}~\bibnamefont {Matsuda}}, \ and\ \bibinfo {author}
  {\bibfnamefont {T.}~\bibnamefont {Terashima}},\ }\href {\doibase
  10.1063/1.2787969} {\bibfield  {journal} {\bibinfo  {journal} {Applied
  Physics Letters}\ }\textbf {\bibinfo {volume} {91}},\ \bibinfo {eid} {122507}
  (\bibinfo {year} {2007})}\BibitemShut {NoStop}%
\bibitem [{\citenamefont {Foyevtsov}\ \emph {et~al.}(2010)\citenamefont
  {Foyevtsov}, \citenamefont {Reith},\ and\ \citenamefont
  {Huth}}]{Foyevtsov20107064}%
  \BibitemOpen
  \bibfield  {author} {\bibinfo {author} {\bibfnamefont {O.}~\bibnamefont
  {Foyevtsov}}, \bibinfo {author} {\bibfnamefont {H.}~\bibnamefont {Reith}}, \
  and\ \bibinfo {author} {\bibfnamefont {M.}~\bibnamefont {Huth}},\ }\href
  {\doibase DOI: 10.1016/j.tsf.2010.06.028} {\bibfield  {journal} {\bibinfo
  {journal} {Thin Solid Films}\ }\textbf {\bibinfo {volume} {518}},\ \bibinfo
  {pages} {7064 } (\bibinfo {year} {2010})}\BibitemShut {NoStop}%
\bibitem [{\citenamefont {Huth}\ \emph {et~al.}(2009)\citenamefont {Huth},
  \citenamefont {Klingenberger}, \citenamefont {Grimm}, \citenamefont
  {Porrati},\ and\ \citenamefont {Sachser}}]{1367-2630-11-3-033032}%
  \BibitemOpen
  \bibfield  {author} {\bibinfo {author} {\bibfnamefont {M.}~\bibnamefont
  {Huth}}, \bibinfo {author} {\bibfnamefont {D.}~\bibnamefont {Klingenberger}},
  \bibinfo {author} {\bibfnamefont {C.}~\bibnamefont {Grimm}}, \bibinfo
  {author} {\bibfnamefont {F.}~\bibnamefont {Porrati}}, \ and\ \bibinfo
  {author} {\bibfnamefont {R.}~\bibnamefont {Sachser}},\ }\href
  {http://stacks.iop.org/1367-2630/11/i=3/a=033032} {\bibfield  {journal}
  {\bibinfo  {journal} {New Journal of Physics}\ }\textbf {\bibinfo {volume}
  {11}},\ \bibinfo {pages} {033032} (\bibinfo {year} {2009})}\BibitemShut
  {NoStop}%
\bibitem [{\citenamefont {Porrati}\ \emph {et~al.}(2009)\citenamefont
  {Porrati}, \citenamefont {Sachser},\ and\ \citenamefont
  {Huth}}]{0957-4484-20-19-195301}%
  \BibitemOpen
  \bibfield  {author} {\bibinfo {author} {\bibfnamefont {F.}~\bibnamefont
  {Porrati}}, \bibinfo {author} {\bibfnamefont {R.}~\bibnamefont {Sachser}}, \
  and\ \bibinfo {author} {\bibfnamefont {M.}~\bibnamefont {Huth}},\ }\href
  {http://stacks.iop.org/0957-4484/20/i=19/a=195301} {\bibfield  {journal}
  {\bibinfo  {journal} {Nanotechnology}\ }\textbf {\bibinfo {volume} {20}},\
  \bibinfo {pages} {195301} (\bibinfo {year} {2009})}\BibitemShut {NoStop}%
\bibitem [{\citenamefont {Ambegaokar}\ and\ \citenamefont
  {Baratoff}(1963)}]{PhysRevLett.10.486}%
  \BibitemOpen
  \bibfield  {author} {\bibinfo {author} {\bibfnamefont {V.}~\bibnamefont
  {Ambegaokar}}\ and\ \bibinfo {author} {\bibfnamefont {A.}~\bibnamefont
  {Baratoff}},\ }\href {\doibase 10.1103/PhysRevLett.10.486} {\bibfield
  {journal} {\bibinfo  {journal} {Phys. Rev. Lett.}\ }\textbf {\bibinfo
  {volume} {10}},\ \bibinfo {pages} {486} (\bibinfo {year} {1963})},\ \bibinfo
  {note} {ibid.{\bf 11}, 104(E) (1963)}\BibitemShut {NoStop}%
\bibitem [{\citenamefont {Petrovic}\ \emph {et~al.}(2001)\citenamefont
  {Petrovic}, \citenamefont {Pagliuso}, \citenamefont {Hundley}, \citenamefont
  {Movshovich}, \citenamefont {Sarrao}, \citenamefont {Thompson}, \citenamefont
  {Fisk},\ and\ \citenamefont {Monthoux}}]{0953-8984-13-17-103}%
  \BibitemOpen
  \bibfield  {author} {\bibinfo {author} {\bibfnamefont {C.}~\bibnamefont
  {Petrovic}}, \bibinfo {author} {\bibfnamefont {P.~G.}\ \bibnamefont
  {Pagliuso}}, \bibinfo {author} {\bibfnamefont {M.~F.}\ \bibnamefont
  {Hundley}}, \bibinfo {author} {\bibfnamefont {R.}~\bibnamefont {Movshovich}},
  \bibinfo {author} {\bibfnamefont {J.~L.}\ \bibnamefont {Sarrao}}, \bibinfo
  {author} {\bibfnamefont {J.~D.}\ \bibnamefont {Thompson}}, \bibinfo {author}
  {\bibfnamefont {Z.}~\bibnamefont {Fisk}}, \ and\ \bibinfo {author}
  {\bibfnamefont {P.}~\bibnamefont {Monthoux}},\ }\href
  {http://stacks.iop.org/0953-8984/13/i=17/a=103} {\bibfield  {journal}
  {\bibinfo  {journal} {Journal of Physics: Condensed Matter}\ }\textbf
  {\bibinfo {volume} {13}},\ \bibinfo {pages} {L337} (\bibinfo {year}
  {2001})}\BibitemShut {NoStop}%
\bibitem [{\citenamefont {Ambegaokar}\ and\ \citenamefont
  {Halperin}(1969)}]{PhysRevLett.22.1364}%
  \BibitemOpen
  \bibfield  {author} {\bibinfo {author} {\bibfnamefont {V.}~\bibnamefont
  {Ambegaokar}}\ and\ \bibinfo {author} {\bibfnamefont {B.~I.}\ \bibnamefont
  {Halperin}},\ }\href {\doibase 10.1103/PhysRevLett.22.1364} {\bibfield
  {journal} {\bibinfo  {journal} {Phys. Rev. Lett.}\ }\textbf {\bibinfo
  {volume} {22}},\ \bibinfo {pages} {1364} (\bibinfo {year}
  {1969})}\BibitemShut {NoStop}%
\bibitem [{\citenamefont {Likharev}(1979)}]{RevModPhys.51.101}%
  \BibitemOpen
  \bibfield  {author} {\bibinfo {author} {\bibfnamefont {K.~K.}\ \bibnamefont
  {Likharev}},\ }\href {\doibase 10.1103/RevModPhys.51.101} {\bibfield
  {journal} {\bibinfo  {journal} {Rev. Mod. Phys.}\ }\textbf {\bibinfo {volume}
  {51}},\ \bibinfo {pages} {101} (\bibinfo {year} {1979})}\BibitemShut
  {NoStop}%
\bibitem [{\citenamefont {Golubov}\ \emph {et~al.}(2004)\citenamefont
  {Golubov}, \citenamefont {Kupriyanov},\ and\ \citenamefont
  {Il'ichev}}]{RevModPhys.76.411}%
  \BibitemOpen
  \bibfield  {author} {\bibinfo {author} {\bibfnamefont {A.~A.}\ \bibnamefont
  {Golubov}}, \bibinfo {author} {\bibfnamefont {M.~Y.}\ \bibnamefont
  {Kupriyanov}}, \ and\ \bibinfo {author} {\bibfnamefont {E.}~\bibnamefont
  {Il'ichev}},\ }\href {\doibase 10.1103/RevModPhys.76.411} {\bibfield
  {journal} {\bibinfo  {journal} {Rev. Mod. Phys.}\ }\textbf {\bibinfo {volume}
  {76}},\ \bibinfo {pages} {411} (\bibinfo {year} {2004})}\BibitemShut
  {NoStop}%
\bibitem [{\citenamefont {DeBeer-Schmitt}\ \emph {et~al.}(2006)\citenamefont
  {DeBeer-Schmitt}, \citenamefont {Dewhurst}, \citenamefont {Hoogenboom},
  \citenamefont {Petrovic},\ and\ \citenamefont
  {Eskildsen}}]{PhysRevLett.97.127001}%
  \BibitemOpen
  \bibfield  {author} {\bibinfo {author} {\bibfnamefont {L.}~\bibnamefont
  {DeBeer-Schmitt}}, \bibinfo {author} {\bibfnamefont {C.~D.}\ \bibnamefont
  {Dewhurst}}, \bibinfo {author} {\bibfnamefont {B.~W.}\ \bibnamefont
  {Hoogenboom}}, \bibinfo {author} {\bibfnamefont {C.}~\bibnamefont
  {Petrovic}}, \ and\ \bibinfo {author} {\bibfnamefont {M.~R.}\ \bibnamefont
  {Eskildsen}},\ }\href {\doibase 10.1103/PhysRevLett.97.127001} {\bibfield
  {journal} {\bibinfo  {journal} {Phys. Rev. Lett.}\ }\textbf {\bibinfo
  {volume} {97}},\ \bibinfo {pages} {127001} (\bibinfo {year}
  {2006})}\BibitemShut {NoStop}%
\bibitem [{\citenamefont {Aslamazov}\ and\ \citenamefont
  {Larkin}(1975)}]{AL.1975}%
  \BibitemOpen
  \bibfield  {author} {\bibinfo {author} {\bibfnamefont {L.~G.}\ \bibnamefont
  {Aslamazov}}\ and\ \bibinfo {author} {\bibfnamefont {A.~I.}\ \bibnamefont
  {Larkin}},\ }\href@noop {} {\bibfield  {journal} {\bibinfo  {journal} {Sov.
  Phys.-JETP}\ }\textbf {\bibinfo {volume} {41}},\ \bibinfo {pages} {381}
  (\bibinfo {year} {1975})}\BibitemShut {NoStop}%
\bibitem [{\citenamefont {Kulik}\ and\ \citenamefont
  {Omelyanchuk}(1975)}]{KO.1975}%
  \BibitemOpen
  \bibfield  {author} {\bibinfo {author} {\bibfnamefont {I.~O.}\ \bibnamefont
  {Kulik}}\ and\ \bibinfo {author} {\bibfnamefont {A.~N.}\ \bibnamefont
  {Omelyanchuk}},\ }\href@noop {} {\bibfield  {journal} {\bibinfo  {journal}
  {Sov. Phys.-JETP}\ }\textbf {\bibinfo {volume} {21}},\ \bibinfo {pages} {96}
  (\bibinfo {year} {1975})},\ \bibinfo {note} {[JETP Lett. {\bf 21}, 96
  (1975)]}\BibitemShut {NoStop}%
\bibitem [{\citenamefont {Kulik}\ and\ \citenamefont
  {Omelyanchuk}(1977)}]{KO.1977}%
  \BibitemOpen
  \bibfield  {author} {\bibinfo {author} {\bibfnamefont {I.~O.}\ \bibnamefont
  {Kulik}}\ and\ \bibinfo {author} {\bibfnamefont {A.~N.}\ \bibnamefont
  {Omelyanchuk}},\ }\href@noop {} {\bibfield  {journal} {\bibinfo  {journal}
  {Fiz. Nizk. Temp.}\ }\textbf {\bibinfo {volume} {3}},\ \bibinfo {pages} {945}
  (\bibinfo {year} {1977})},\ \bibinfo {note} {[Sov. J. Low Temp. Phys. {\bf
  3}, 459 (1977)]}\BibitemShut {NoStop}%
\bibitem [{\citenamefont {Ishii}(1970)}]{PTP.44.1525}%
  \BibitemOpen
  \bibfield  {author} {\bibinfo {author} {\bibfnamefont {C.}~\bibnamefont
  {Ishii}},\ }\href {\doibase 10.1143/PTP.44.1525} {\bibfield  {journal}
  {\bibinfo  {journal} {Progress of Theoretical Physics}\ }\textbf {\bibinfo
  {volume} {44}},\ \bibinfo {pages} {1525} (\bibinfo {year}
  {1970})}\BibitemShut {NoStop}%
\bibitem [{\citenamefont {\"{O}zcan}\ \emph {et~al.}(2003)\citenamefont
  {\"{O}zcan}, \citenamefont {Broun}, \citenamefont {Morgan}, \citenamefont
  {Haselwimmer}, \citenamefont {Sarrao}, \citenamefont {Kamal}, \citenamefont
  {Bidinosti}, \citenamefont {Turner}, \citenamefont {Raudsepp},\ and\
  \citenamefont {Waldram}}]{0295-5075-62-3-412}%
  \BibitemOpen
  \bibfield  {author} {\bibinfo {author} {\bibfnamefont {S.}~\bibnamefont
  {\"{O}zcan}}, \bibinfo {author} {\bibfnamefont {D.~M.}\ \bibnamefont
  {Broun}}, \bibinfo {author} {\bibfnamefont {B.}~\bibnamefont {Morgan}},
  \bibinfo {author} {\bibfnamefont {R.~K.~W.}\ \bibnamefont {Haselwimmer}},
  \bibinfo {author} {\bibfnamefont {J.~L.}\ \bibnamefont {Sarrao}}, \bibinfo
  {author} {\bibfnamefont {S.}~\bibnamefont {Kamal}}, \bibinfo {author}
  {\bibfnamefont {C.~P.}\ \bibnamefont {Bidinosti}}, \bibinfo {author}
  {\bibfnamefont {P.~J.}\ \bibnamefont {Turner}}, \bibinfo {author}
  {\bibfnamefont {M.}~\bibnamefont {Raudsepp}}, \ and\ \bibinfo {author}
  {\bibfnamefont {J.~R.}\ \bibnamefont {Waldram}},\ }\href
  {http://stacks.iop.org/0295-5075/62/i=3/a=412} {\bibfield  {journal}
  {\bibinfo  {journal} {EPL (Europhysics Letters)}\ }\textbf {\bibinfo {volume}
  {62}},\ \bibinfo {pages} {412} (\bibinfo {year} {2003})}\BibitemShut
  {NoStop}%
\bibitem [{\citenamefont {Bardeen}\ and\ \citenamefont
  {Stephen}(1965)}]{PhysRev.140.A1197}%
  \BibitemOpen
  \bibfield  {author} {\bibinfo {author} {\bibfnamefont {J.}~\bibnamefont
  {Bardeen}}\ and\ \bibinfo {author} {\bibfnamefont {M.~J.}\ \bibnamefont
  {Stephen}},\ }\href {\doibase 10.1103/PhysRev.140.A1197} {\bibfield
  {journal} {\bibinfo  {journal} {Phys. Rev.}\ }\textbf {\bibinfo {volume}
  {140}},\ \bibinfo {pages} {A1197} (\bibinfo {year} {1965})}\BibitemShut
  {NoStop}%
\bibitem [{\citenamefont {Gumann}\ \emph {et~al.}(2007)\citenamefont {Gumann},
  \citenamefont {Dahm},\ and\ \citenamefont {Schopohl}}]{PhysRevB.76.064529}%
  \BibitemOpen
  \bibfield  {author} {\bibinfo {author} {\bibfnamefont {A.}~\bibnamefont
  {Gumann}}, \bibinfo {author} {\bibfnamefont {T.}~\bibnamefont {Dahm}}, \ and\
  \bibinfo {author} {\bibfnamefont {N.}~\bibnamefont {Schopohl}},\ }\href
  {\doibase 10.1103/PhysRevB.76.064529} {\bibfield  {journal} {\bibinfo
  {journal} {Phys. Rev. B}\ }\textbf {\bibinfo {volume} {76}},\ \bibinfo
  {pages} {064529} (\bibinfo {year} {2007})}\BibitemShut {NoStop}%
\bibitem [{\citenamefont {Gumann}\ and\ \citenamefont
  {Schopohl}(2009)}]{PhysRevB.79.144505}%
  \BibitemOpen
  \bibfield  {author} {\bibinfo {author} {\bibfnamefont {A.}~\bibnamefont
  {Gumann}}\ and\ \bibinfo {author} {\bibfnamefont {N.}~\bibnamefont
  {Schopohl}},\ }\href {\doibase 10.1103/PhysRevB.79.144505} {\bibfield
  {journal} {\bibinfo  {journal} {Phys. Rev. B}\ }\textbf {\bibinfo {volume}
  {79}},\ \bibinfo {pages} {144505} (\bibinfo {year} {2009})}\BibitemShut
  {NoStop}%
\end{thebibliography}


%

\end{document}